\documentclass{Earthzine}

\newcommand{\EZyear}{2014}
\newcommand{\EZvolume}{7}

\newcommand{\EZissue}{2}
\newcommand{\EZissueText}{$2^{nd}$ quarter theme. Geospatial Semantic Array Programming}

\EZcomment{}
\newcommand{\EZpage}{877975}

\title{Applying Geospatial Semantic Array Programming for a Reproducible Set of Bioclimatic Indices in Europe}

\newcommand{\EZsearch}{Applying+Geospatial+Semantic+Array+Programming+for+a+Reproducible+Set+of+Bioclimatic+Indices+in+Europe}
\EZregisterTitle

\EZauthor{1}{Giovanni}{G.}{Caudullo}
\EZaffiliation{1}{European Commission, Joint Research Centre, Institute for Environment and Sustainability, \vspace{-2mm}

Forest Resources and Climate Unit, Via E. Fermi 2749, I-21027 Ispra, VA, Italy
}

\EZregisterAuthors

\EZregisterAbstract{Bioclimate-driven regression analysis is a widely used approach for modelling ecological niches and zonation. Although the bioclimatic complexity of the European continent is high, a particular combination of 12 climatic and topographic covariates was recently found able to reliably reproduce the ecological zoning of the Food and Agriculture Organization of the United Nations (FAO) for forest resources assessment at pan-European scale, generating the first fuzzy similarity map of FAO ecozones in Europe. The reproducible procedure followed to derive this collection of bioclimatic indices is now presented. It required an integration of data-transformation modules (D-TM) using geospatial tools such as Geographic Information System (GIS) software, and array-based mathematical implementation such as semantic array programming (SemAP). Base variables, intermediate and final covariates are described and semantically defined by providing the workflow of D-TMs and the mathematical formulation following the SemAP notation. Source layers to derive base variables were extracted by exclusively relying on global-scale public open geodata in order for the same set of bioclimatic covariates to be reproducible in any region worldwide. In particular, two freely available datasets were exploited for temperature and precipitation (WorldClim) and elevation (Global Multi-resolution Terrain Elevation Data). The working extent covers the European continent to the Urals with a resolution of 30 arc-second. The proposed set of bioclimatic covariates will be made available as open data in the European Forest Data Centre (EFDAC). The forthcoming complete set of D-TM codelets will enable the 12 covariates to be easily reproduced and expanded through free software.}

\begin{document}

\EZmaketitle

\section*{Introduction}

The integrated use of geospatial tools and regression analyses (from basic approaches such as distance weighting, polynomial interpolation and regression, up to more complex nonlinear regression techniques) is widely expanding for modelling habitat suitability, ecological niches and zonation starting from field observations (i.e., presence-only or presence/absence of species) \cite{Store_Jokimaki_2003}\cite{Brotons_etal_2004}\cite{Hirzel_Le_Lay_2008}. The selection of predictor variables is an important factor for developing species distribution models \cite{Araujo_Guisan_2006}. In this work a set of twelve bioclimatic covariates are derived from publicly-available datasets and proposed for ecological niche modelling and zonation. This particular combination of climatic and topographic indices has been elaborated by de Rigo et al. \cite{de_Rigo_etal_Continental2013} to reliably reproduce the ecological zoning of the Food and Agriculture Organization of the United Nations (FAO) for forest resources assessment at pan-European scale, generating the first fuzzy similarity map of FAO ecozones in Europe. The original layer of FAO ecological zones was produced on the basis of bioclimatic variables and potential vegetation and was subject to expert-based refinement \cite{FAO_2012}. The final product is distributed as a crisp (i.e., non-fuzzy) vector file. The application of a model based on the relative distance similarity (RDS) technique allowed each ecological zone to be compared with climatically and geographically analogous grid-cells, reducing boundary artefacts and modelling a fuzzy raster map at 1 square kilometer. The analogous approach has been used in Ciscar et al. \cite{Ciscar_etal_2014} performing the RDS method to estimate the impacts of habitat suitability change of a forest tree species over Europe for different future climate scenarios. The maximum suitability maps have been computed on the basis of field survey datasets of tree species presence/absence, modelled with a set of bioclimatic and geographic indicators.

This proposed set of predictors has been selected on the basis of ecological and bio-geographical knowledge and adapted to the European climate. In this paper, these covariates are presented with emphasis on their definitions, technical details and modelling procedures. The approach of reproducibility has been followed in all processes in order to allow these covariates to be easily re- used and/or modified for further ecological regression analyses of niche/zoning modelling. 

\section*{Datasets}
The datasets used for extracting base variables are freely available as Geographic Information System (GIS) layers at global scale. 

\href{http://www.worldclim.org}{\textbf{WorldClime}}  v1.4: is a set of global climate layers with a spatial resolution of 30 arc-second, which consists of four climate variables: precipitation and mean, minimum, maximum temperature. Data were gathered from a variety of resources, covering as a current baseline the period from 1950 to 2000. Future scenarios developed with different models also are available. There are 12 layers for every variable, representing its 50-years monthly average values \cite{Hijmans_etal_2005}.

\href{https://lta.cr.usgs.gov/GMTED2010}{\textbf{Global Multi-resolution Terrain Elevation Data 2010}} (GMTED): is an enhanced elevation model developed by the U.S. Geological Survey and the National Geospatial-Intelligence Agency at global scale. The product, available in different spatial resolutions up to 7.5 arc-second, is based on data derived from 11 raster-based elevation sources, and aggregated in six products: minimum, maximum, mean and median elevation, standard deviation of elevation, systematic subsample, breakline emphasis \cite{Danielson_Gesch_2011}.

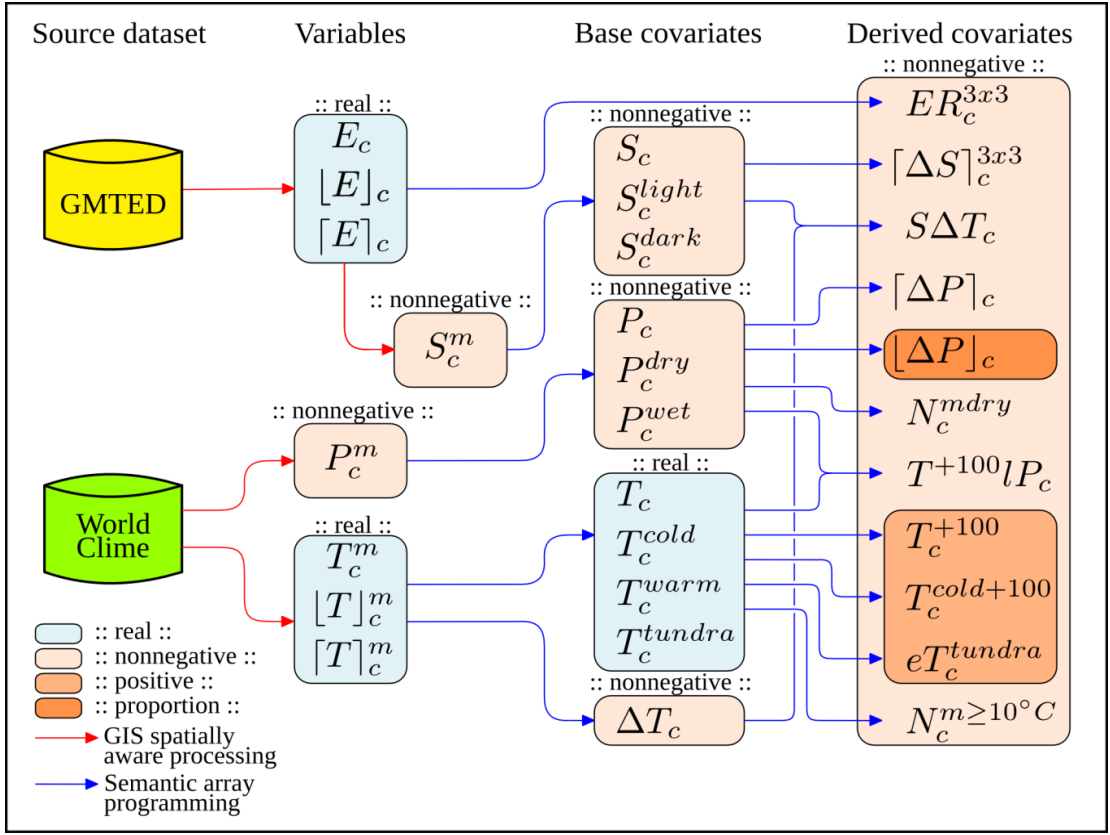
\fullEZimage{Figure1}{Figure 1. Process modelling workflow: the arrows represent the data transformation modules to derive the indices, which are grouped and coloured according to their semantic constrains.}

\section*{Methods}

The process for modelling the final covariates required different data-transformation modules (D-TMs) integrating geospatial tools and array-based mathematical implementations such as semantic array programming (i.e., GeoSemAP) \cite{de_Rigo_etal_Toward2013}.
The workflow in Figure 1 shows D-TMs from source dataset to bioclimatic covariates. The colours of arrows define the transformation methods, while the colours of the grouping subsets of indices define the semantics of the outputs and intermediate layers. GIS technologies (red arrows) have been used principally for extracting data from source layers and preparing base covariates as a \href{http://trac.osgeo.org/geotiff}{GeoTIFF} raster. In particular, Python scripting \cite{Van_Rossum_Drake_2011} with \href{http://www.gdal.org}{GDAL/OGR} libraries \cite{Warmerdam_2008}\cite{GDAL_2014} and \href{http://grass.osgeo.org}{Grass GIS} v.7 libraries \cite{Neteler_etal_2012} have been used to operate with spatial layers. Data-transformation for converting raster maps into semantic array and SemAP procedure (blue arrows) have been based on GNU Octave/MATLAB programming language \cite{Eaton_2012} with \href{http://mastrave.org}{Mastrave} library \cite{de_Rigo_2012} and on Python language with \href{http://www.scipy.org}{SciPy} library \cite{Jones_etal_2001}. Indices are grouped in modules and colored according to their semantic constraints, which are exemplified with the notation \href{http://mastrave.org/doc/mtv_m/check_is}{::constrains::}.
The working spatial environment is established to cover Europe completely to the Urals (north 72 degrees, south 28 degrees, east 75 degrees, west -25 degrees). The adopted projection is WGS84 with a spatial resolution of 30 arc-second, operating with GeoTIFF raster files with a dimension of approximately 63 million pixels as floating point values. Variables are defined in each spatial grid c and for each month m. In Table 1, both variables and covariates are described with relative symbols, descriptions and reference equations. The temperature T is expressed in degree Celsius + 100, avoiding negative values. Temperature without symbols is defined as the average value of the month, while symbol $\lceil \;\rceil$ defines the \textit{maximum} of the month and the symbol $\lfloor \;\rfloor$ the \textit{minimum}. The precipitation P is in millimeters and values state the total sum of monthly precipitation. The elevation E without symbols is the average value inside the 1 km grid measured in meters. The solar irradiation S is expressed in watt-hour per square meter ($\text{Whm}^{\text{-2}}$). Quantities averaged with a spatial moving window of 3 by 3 km are denoted with $^{\text{3x3}}$. The definition of variable attributes \textit{minimum}, \textit{average} and \textit{maximum} guarantees them to be semantically sortable ensuring \href{http://mastrave.org/doc/mtv_m/check_is#SAP_nonnegative}{::nonnegative::} values for the difference between any pair of attributes in ascending order, thus eliminating potential inconsistencies in the original data.

\newcommand{\OneSquareKm}{\ensuremath{1\,km\!\times\!1\,km}}
\newcommand{\ThreeSquareKm}{\ensuremath{3\,km\!\times\!3\,km}} 
\makeatletter
\newcommand*\tablesize{%
  \@setfontsize\tablesize{9}{11.0}%
}
\makeatother
\newcommand{\Celsius}{\ensuremath{{}^{\,\circ\!}C}}
\newcolumntype{x}[1]{>{\centering\hspace{0pt}}p{#1}}
\nohyphens{
\begin{table}\vspace{-0mm}
\caption{{Table 1. List of the computed variables and covariates with relative descriptions and equations.}}
\centerline{\tablesize 
\begin{tabular}{p{2.4cm} p{6.92cm} p{3.5cm} }
\hline\noalign{\smallskip}
\begin{tabular}[c]{c}
Variable \\[-1mm]
code \\[0.3mm]
\end{tabular} & \multicolumn{2}{c}{Description of variable} \\
\hline
\noalign{\smallskip}
$E_c$                    & Mean elevation & \\
${\lfloor E\rfloor}_c$   & Minimum elevation & \\
${\lceil E\rceil}_c$     & Maximum elevation & \\
$S_c^m$                  & \multicolumn{2}{l}{Total solar irradiation in the central day of the month m} \\
$P_c^m$                  & Total precipitation of the month m & \\
$T_c^m$                  & Average temperature of the month m & \\
${\lfloor T\rfloor}_c^m$ & Minimum temperature of the month m & \\
${\lceil T\rceil}_c^m$   & Maximum temperature of the month m & \\
%
%
\noalign{\smallskip}
\hline\noalign{\vspace{1.5mm}}
\hline\noalign{\smallskip}
\begin{tabular}[c]{c}
Base \\[-1mm]
covariate \\[0.3mm]
\end{tabular} & \multicolumn{1}{c}{Description of the base covariate} & Reference equation\\
\hline
\noalign{\smallskip}
$S_c^{\text{light}}$  & Total potential solar irr. of the 6 lighter months & \\
$S_c^{\text{dark}}$   & Total potential solar irr. of the 6 darker  months & \\
$P_c$                 & Annual total precipitation  & $\sum_m P_c^m$\\
$P_c^{\text{dry}}$    & Total precipitation of the driest month  & $\min_m P_c^m$\\
$P_c^{\text{wet}}$    & Total precipitation of the wettest month & $\max_m P_c^m$\\
$T_c$                 & Annual average temperature & $\sum_m T_c^m$ \\
$T_c^{\text{cold}}$   & Average temperature of the coldest month & $\min_m T_c^m$\\
$T_c^{\text{warm}}$   & Average temperature of the warmest month & $\max_m T_c^m$\\
$T_c^{\text{tundra}}$ & Nordenski\"old index & $T_c^{\text{warm}} + 0.1 T_c^{\text{cold}} -9 $\\
$\Delta T_c$ & Mean of monthly temperature range & $1/12 \sum_{m} {\lceil T \rceil}_c^m \!-\! {\lfloor T \rfloor}_c^m $ \\
\noalign{\smallskip}
\hline
\end{tabular}} 
%
%
\smallskip
\centerline{\tablesize 
\begin{tabular}{p{0.3cm} p{1.65cm} p{6.92cm} p{3.5cm}}
\hline\noalign{\smallskip}
\multicolumn{2}{c}{\begin{tabular}[c]{c}
Derived \\[-1mm]
covariates \\[0.3mm]
\end{tabular}} & \multicolumn{1}{c}{Description of derived covariate} & Reference equation\\
\hline
\noalign{\smallskip}
1.& $\text{\textit{ER}}_c^{3\times 3}$ & \ThreeSquareKm{} average elevation range & $({\lceil E \rceil}_c \!-\! {\lfloor E \rfloor}_c)^{3\times 3} $\\[1mm]
2.& $S_c$ & Annual potential solar irradiation & $S_c^{\text{light}} + S_c^{\text{dark}}$ \\[0.5mm]
3.& ${\lceil\Delta S\rceil}_c^{3\times 3}$ & \ThreeSquareKm{} seasonal variation of pot. solar irr. (normalised by semestral minimum) & $\displaystyle \frac{S_c^{\text{light},3\times 3} - S_c^{\text{dark},3\times 3}}{S_c^{\text{dark},3\times 3}}$\\
4.&$S\Delta T_c$ & Potential irradiation by montly temp. range (simplified avg. monthly solar entransy flux) & $S_c \cdot \Delta T_c $\\
5.&${\lceil\Delta P\rceil}_c$   & Seasonal variation of monthly precipitation (normalised by minimum: driest month) & $\displaystyle \frac{P_c^{\text{wet}} - P_c^{\text{dry}}}{P_c^{\text{dry}}}$\\
6.& ${\lfloor\Delta P\rfloor}_c$ & Seasonal variation of monthly precipitation (normalised by maximum: wettest month) & $\displaystyle \frac{P_c^{\text{wet}} - P_c^{\text{dry}}}{P_c^{\text{wet}}}$\\
7.&$N_c^{m \text{dry}}$ & Number of dry months & $\sum_m [P_c^m \le 2 T_c^m ]$\\[0.7mm]
8.&$T^{+\!100}lP_c$ & Shifted temperature per rain magnitude order & $T_c^{+\!100} / \log_{10}(\, P_c +1\,)$ \\[0.5mm]
9.&$T_c^{+\!100}$ & Shifted average annual temperature & $T_c + 100\Celsius{}$ \\[0.5mm]
10.&$T_c^{\text{cold} +\!100}$ & Shifted average temp. of the coldest month & $T_c^{\text{cold}} + 100\Celsius{}$ \\[0.5mm]
11.&$eT_c^{\text{tundra}}$ & Nordenski\"old exponential index & $\exp(\, T_c^{\text{tundra}} \,)$ \\[0.5mm]
12.&$N_c^{m \ge 10\Celsius{}}$ & Number of months with $T_c \ge 10\Celsius{} $ & $\sum_m [ T_c^m \ge 10 ]$\\
\noalign{\smallskip} 
\hline
\end{tabular}}  
\label{tab:biocovs}\vspace{-4mm}
\end{table}
}


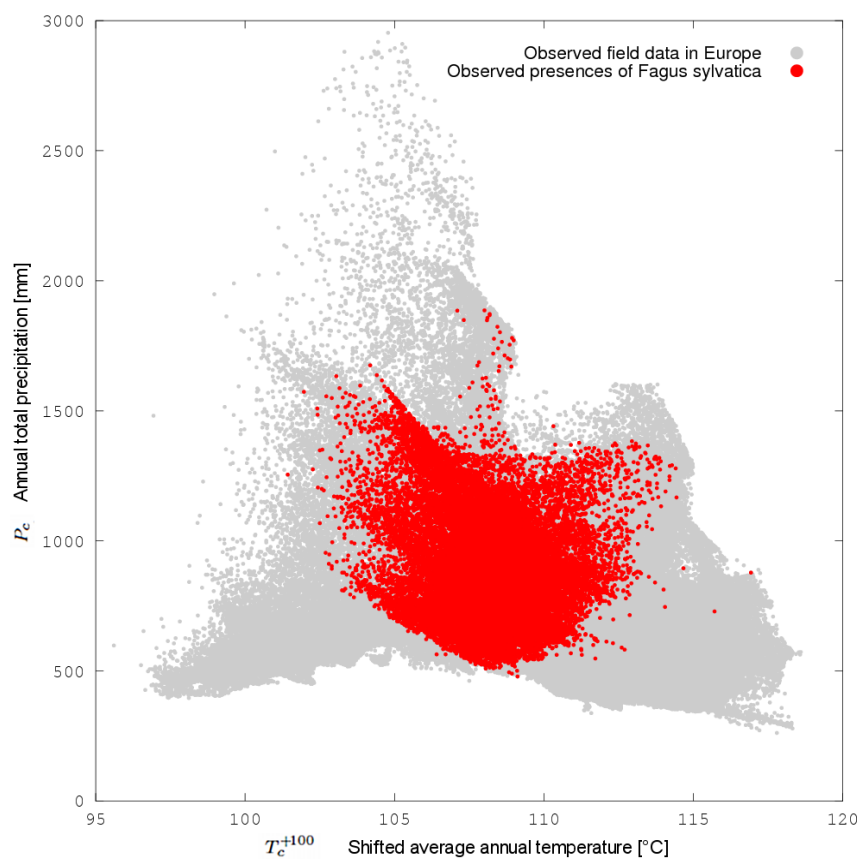
\customEZimage{Figure2}{Figure 2. Ecological niche diagram of European beech ({\upshape Fagus sylvatica}) comparing the annual total precipitation and annual average temperature (shifted +100 degrees Celsius): the observed presence of the species (red dots) is highlighted over nearly 250,000 sample points (gray dots).}{0.8}{0.8}

Temperature and precipitation variables (from WorldClime raster dataset) have “No Data” values covering water pixels, converted to “not a number” (NaN) values on semantic arrays. NaN creates a water mask which propagates on array computations up to derived covariates.
The Tundra covariate is identified by the equation of Nordenskiold and Mecking  \cite{Nordenskiold_Mecking_1928}. The average temperature of coldest month and the number of dry months and months with a temperature higher than 10 degrees Celsius are defined by FAO for the ecological zones \cite{FAO_2012}. The combination of potential solar irradiation by temperature range is a simplification of the solar entransy flux \cite{Guo_etal_2007}\cite{Xu_2011}.
Covariates counting the number of months were calculated tacking into account the temperature variation within the 1x1 km spatial grid due to difference in elevation adopting the lapse rate. Covariates derived from solar irradiation required heavy computations performed by Grass GIS v.7 software. The potential solar irradiation was selected instead of the actual one due to its stability under different climate change scenarios. It was modelled starting from the average of the GMTED with a higher spatial resolution at 7.5 arc-second, gaining raster dimension of 273 million pixels. It is proven that the solar irradiation derived from a digital elevation model at 30 arc-second is less accurate compared with the aggregation to 30 arc-second of solar irradiation derived from a 7.5 arc-second. The average elevation was calculated using \href{http://grass.osgeo.org/grass70/manuals/r.sun.html}{r.sun} Grass module \cite{Hofierka_etal_2007}, combining the shadowing effects of \href{http://grass.osgeo.org/grass70/manuals/r.horizon.html}{r.horizon} module derived every 5 degrees \cite{Huld_etal_2007}. The atmospheric turbidity and albedo effects were not modelled, thus solar energy layer is a proxy of the potential one and essentially dependent on the aspect, latitude and orography-induced patterns of shadows. Twelve maps of solar irradiation were produced for the central days of each month. The monthly aggregation was computed by integrating data of grid maps using the trapezoidal rule. Thereafter the solar irradiation covariates were harmonized to 30 arc-second through a spatial mean aggregation.

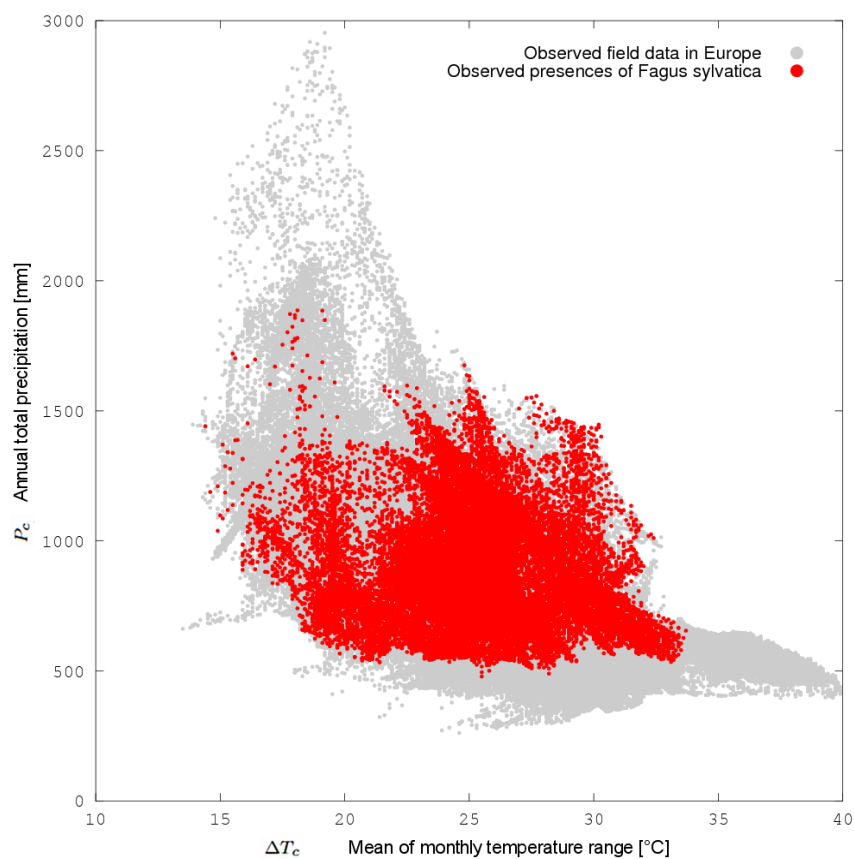
\customEZimage{Figure3}{Figure 3. Ecological niche diagram of European beech ({\upshape Fagus sylvatica}) comparing the annual total precipitation and mean of monthly temperature range: the observed presence of the species (red dots) is highlighted over nearly 250,000 sample points (gray dots).}{0.8}{0.8}

\section*{Results}

All described covariates will be visualized and downloadable on the \href{http://efdac.jrc.ec.europa.eu}{European Forest Data Centre} of the \href{http://fise.ec.europa.eu}{Forest Information System for Europe} (EFDAC-FISE). The modelling procedures have been performed with free software packages which allow a complete reproducibility of D-TMs. 
This set of 12 bioclimatic covariates, selected and elaborated for the first time to reproduce European forest ecological zones, is a combination of regression analysis predictors suitable for biological niche modelling related principally to forest resources at European scale (e.g., forest type and species suitability maps \cite{Rehfeldt_etal_1999}\cite{Rupprecht_etal_2011}\cite{Conlisk_etal_2012} and forest pest outbreak predictions \cite{Aukema_etal_2008}\cite{SobekSwant_etal_2012}\cite{de_Rigo_etal_2014}). The same proposed set of predictors will be processed with the RDS regression method for the European Atlas of Forest Tree Species (in EFDAC-FISE), a publication where habitat suitability maps of main European tree species also will be presented. The codelets of described procedures were organized in modules corresponding to the D-TMs, enabling the complete set of covariate raster maps to be reproduced from different data sources and at different scales (technical details on de Rigo,  Caudullo) \cite{de_Rigo_Caudullo_inprep}. The forthcoming codelets also will be available on the EFDAC-FISE website.

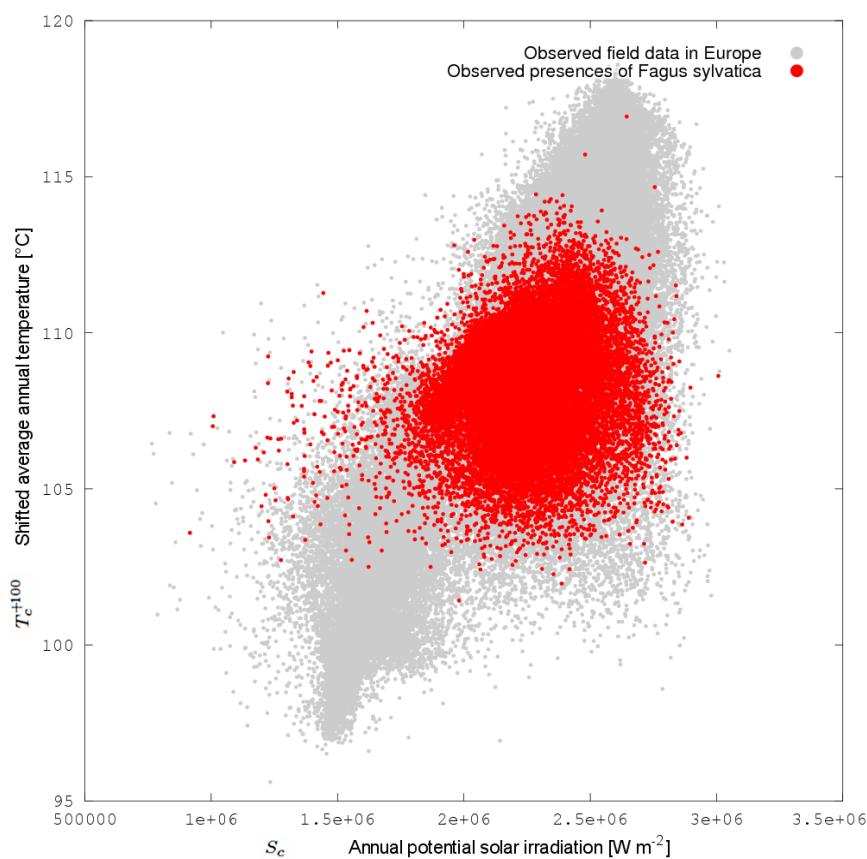
\customEZimage{Figure4}{Figure 4. Ecological niche diagram of European beech ({\upshape Fagus sylvatica}) comparing annual average temperature (shifted +100 degrees Celsius) and the annual potential solar irradiation: the observed presence of the species (red dots) is highlighted over nearly 250,000 sample points (gray dots).}{0.8}{0.8}

In addition to geographic maps, covariates may be represented two by two as diagrams plotting values on the x and y axis within the study area. This family of charts can be used to analyze correlations between variables, and show ecological niche patterns when a certain species presence is highlighted.
To exemplify this possible usage, the realized ecological niche of European beech (\textit{Fagus sylvatica}) has been computed using the EFDAC-FISE datasets, part of them deriving from a harmonization effort of the European National Forest Inventories. The data refer to observed presences of the tree species. In Figure 2, nearly 250,000 sample points are represented for annual precipitation and annual average temperature (gray dots). The presence of European beech is large (red dots) and centered in the middle of the observation cloud, confirming its common and widespread distribution in temperate forests  \cite{Packham_etal_2012}.
Figure 3 shows the relation between precipitation and the annual range in temperature. Here it is more evident that the current presence of European beech is limited by a minimum precipitation of about 500 mm and a maximum of 2,000 mm. Annual range in temperature gives a measure of the continentality of the climate. Beech is clearly absent above a range of 33-34 degrees Celsius in temperature range, confirming that this tree has a realized niche characterized by more oceanic climates \cite{Packham_etal_2012}.
Solar irradiation and annual average temperature covariates show a positive correlation (Figure 4), since ground temperature depends principally on solar energy. The observation cloud shows a linear growing trend. On the other hand, beech presence does not present crisp boundaries on the ecological niche modelled with these two covariates. Isolated samples are still present on the cloud edges, in particular where solar irradiation has lower values, without evident thresholds.

\section*{Author Bio}
\underline{Giovanni Caudullo} holds a degree in Forestry and Environmental Science from the Università degli Studi di Padova. He has since worked for six years as a freelance environmental and GIS consultant for private companies and public authorities. He is currently working at the European Commission’s Joint Research Centre in the Forest Resources and Climate Unit as senior GIS analyst, operating with spatial and temporal indicators from local to Europe-wide scale and developing procedures for automatic data processing.

\nohyphens{

}

\end{document}